# Field-dependent surface resistance for superconducting niobium accelerating cavities - condensed overview of weak superconducting defect model

Wolfgang Weingarten[1]
CERN
Esplanade des Particules 1
P.O. Box
1211 Geneva 23
Switzerland

**Abstract**

Small (compared to coherence length) weak superconducting defects when located at the surface, combined with the proximity and percolation effects, are claimed responsible for various observations with superconducting rf accelerating cavities with non-constant Q-value, such as "Q-slope" and "Q-drop", the role of temperature, the result of "nitrogen doping" and its relation to the free path, and the influence of the static external magnetic field. The Ginzburg-Landau equations are used to confirm the results.

## 1. Introduction

This publication is motivated by former studies on the superconducting (sc) rf system for the LEP and LHC storage rings at CERN. There was a long discussion about the possible choice of technologies, either to build the sc rf cavities from niobium sheet or from copper with a thin niobium coating (Nb/Cu) on top. For LEP, rf cavities made from both technologies were installed, although the bulk of rf cavities was made from Nb/Cu, whereas for the LHC Nb/Cu cavities were used. This decision in favour of the Nb/Cu technology was justified by the lower performance requirements for LEP and LHC in terms of maximum gradient. The Nb/Cu cavities showed a stronger increase of the rf losses with the accelerating gradient than the niobium sheet cavities. Therefore, the latter ones found their application in fairly all recent linear accelerators, such as energy recovery and recirculating machines, etc., as well as in studies on future sc linear colliders (ILC). The reason for this is that, for cost reasons, the overall length of a linear accelerator must be kept as short as possible, since particles pass through the device only once or a few times, whereas in storage rings the accelerating gap is continuously traversed.

However, Nb/Cu technology showed several advantages over niobium sheet technology: lower sensitivity to the ambient magnetic field, better tolerance to lossy defects due to the greater thermal stability of the thin niobium layer supported by copper with high thermal conductivity, and last but not least, lower manufacturing costs. Consequently, there was a strong motivation to understand why the rf losses of the Nb/Cu cavities increased faster with the acceleration gradient than those of the niobium sheet cavities. After a long period of research, this drawback was mitigated but not completely eliminated [1, 2].

The present paper is an attempt to better understand the problem of increased losses in Nb/Cu cavities in particular, but also in niobium sheet cavities. In the published literature several explanations have been published, [3] being the most recent, [4]. These are, for example, suppression of the energy gap or the modification of the density of states with increasing rf field. These explanations are not contested but assumed to be negligibly small compared to the model described in this paper.

To expand on this argument, reference should be made to a measurement of an Nb accelerating cavity where a very high Q-value ($7 \cdot 10^{10}$ at 1.6 K) was relatively constant up to a maximum field strength of about 190 mT [5]. The Q-value at low field and 1.3 K was $1 \cdot 10^{11}$. The authors assume that this cavity had an almost defect-free surface as a result of a number of fortunate circumstances during preparation. Conversely, a field-dependent Q value as being caused by defects is presented in this paper complementing the intrinsic factors as mentioned before.

The following arguments and conclusions are not new in themselves, but have been published in various places (except section 7). Some of them have more the character of an assertion of a speculative nature, although substantiated. On the other hand, these claims gain weight when identical reasoning

---

[1] The author, retired, was with the Department of Beams, CERN, 1211 Geneva, Switzerland (e-mail: wolfgangweingarten@t-online.de).

from other observations summarized in this paper is considered.

The results can be traced back to a common explanation, namely small (compared to the coherence length) weak sc defects when located at the current carrying surface of the sc cavity, in conjunction with the proximity and percolation effects [6, 7].

In the present work, the arguments and observations concerning this explanation are summarized one by one. Only the most important results published so far are given here; for a deeper insight into the physics, the original papers should be consulted. Some associated figures illustrate the arguments.

Table A-1 (Appendix) lists additional information on the symbols used in this text.

## 2. Surface energy balance

There is a gain in diamagnetic (surface) energy at the expense of the loss of condensation energy, when a sufficiently small weak sc defect on the surface transforms into a normal conducting (nc) defect under the action of the rf magnetic field B. This effect is illustrated by the commonly plotted quality factor $Q$ as a function of magnetic field $B$, as shown in Fig. 1, based on a closed form equation, eq. 1.c. More details with regard to eq. 1.c are presented in the appendix A-1.

The data are obtained for a fine grain niobium monocell cavity at 1300 MHz and 2 K [8]. The continuous lines represent the results from fitting the data with the Ginzburg-Landau parameter κ = 1.5, 1.3 and 0.9 (from bottom to top) [6].

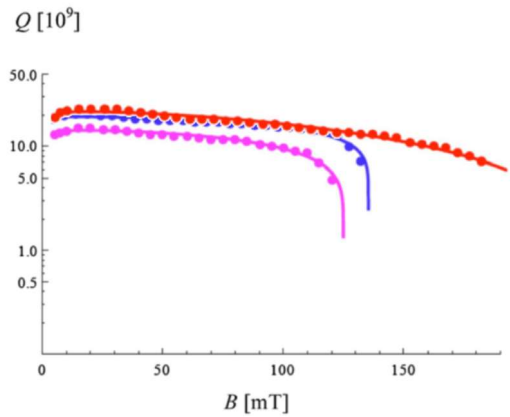

Fig. 1. Typical comparison of measured and fitted individual data: The Q(B) curves were obtained for a 1300 MHz single cell cavity at 2.0 K made of fine grain niobium.

The plot shows the commonly observed dependence of the Q-value on the magnetic field B, usually referred to as "Q-slope" and "Q-drop". From the Q-value the surface resistance $R_s$ can be extracted by $R_s = G/Q$, where G is a constant that depends on the rf field pattern alone. $R_s$, in turn, is considered to be the sum of the residual surface resistance $R_{res}$, which is rf-field and temperature independent, and all other contributions, such as a temperature-dependent term often referred to as the "BCS surface resistance" $R_{BCS}$, and a field-dependent term $R_{s,fd}$,

$$R = R_{res} + R_{BCS}(\omega, T) + R_{s,fd}(B, \omega, T) \quad (1)$$

with

$$R_{BCS}(\omega, T) = \frac{1}{2} \cdot \mu_0^2 \omega^2 \lambda^3 s_{Nb} \cdot \frac{\Delta}{k_B T} \ln\left(\frac{\Delta}{\hbar \omega}\right) e^{-\Delta/T}$$

(1.a)

$$R_{s,fd}(B, \omega, T) = R_{s,fdb}(B) \cdot R_{s,fdt}(\omega, T) \quad (1.b)$$

$$R_{s,fdb}(B) = (-\kappa^{-2}) \cdot \left\{1 + \frac{\ln\left[1-\kappa^2\left(\frac{B}{B_c}\right)^2\right]}{\kappa^2\left(\frac{B}{B_c}\right)^2}\right\} \quad (1.c)$$

$R_{s,fdt}$ turns out to be quite close to $R_{BCS}$ [6]. The abrupt drop of the Q-value at increased magnetic field (e.g., ~ 0.125 T for κ = 1.5) is also explained, when $\kappa^2(B/B_c)^2$ approaches 1. A more detailed derivation of eq. 1.c can be found in the appendix.

A distinction must be made between the situation when the defect is located at the surface and the situation when it is located in the volume but still within a distance given by the penetration depth and thus exposed to the rf current. When the defect is embedded in the volume, the current flows around it on both sides when it becomes nc. In other words, a loop-like microscopic magnetic field is created with the net result that the magnetic induction in the superconductor does not change: the diamagnetic energy remains unchanged, the energy balance does not become negative, and therefore there is no gain in the energy balance. This is the reason why the nc volume under the influence of the magnetic field B can grow only on the surface and not inside the superconductor. Thus, a precondition for growth is the presence of a weak sc defect at the surface.

## 3. Role of temperature

The outermost surface of the rf cavity is assumed to be an inhomogeneous mixture consisting of Nb and NbO, representative of other impurities. Then there are areas with an intimate neighbourhood of Nb and NbO (called composite) and other areas with less or no content of NbO (called Nb matrix).

First, in the composites plus matrix the proximity effect between the strong superconductor (S, Nb) and the weak superconductor (N, NbO) will act. Let the volume fraction of S be x = $v_S/(v_S+v_N)$. The weak superconductor (NbO) by itself has a critical temperature $T_c$ = 1.34 K, but the neighbourhood of the strong superconductor increases its critical temperature $T_{CNS}$ according to x, as prescribed by the Cooper limit approximation which is used here.

Second, at a sufficiently high temperature, some regions of matrix plus composites become normal conducting, while other regions remain superconducting. Then the question arises to what extent the composites together with the matrix fragment into normal conducting and superconducting sites. In other words, the question is as to what proportion of x, as a function of temperature, sufficiently small normal conducting regions will emerge that will serve as expanding defects as described in the previous section. This is a problem of percolation.

According to the Cooper limit approximation, the relation of the critical temperature $T_{cNS}$ of the composite vs the volume ratio x starts at the critical temperature $T_c$ = 1.34 K for the NbO solely, when the concentration of the S component is zero. The temperature difference $T_{cNS} - T_{c,NbO}$ follows a quasilinear relation with the concentration x [6], Fig. 2.

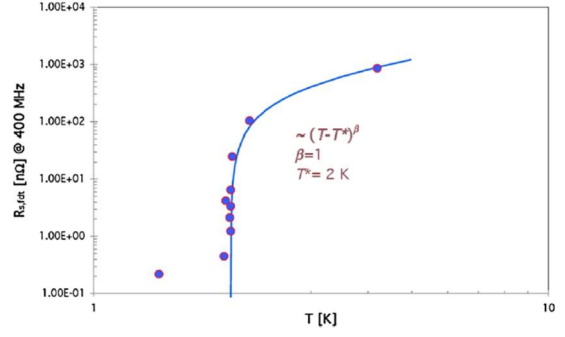

Fig. 3: The average temperature dependent part $R_{s,fdt}$ of the field dependent surface resistance versus the bath temperature T.

It follows as a corollary that long-range connectivity is associated with the generation of isolated sites acting as nucleation centres favouring the entry of magnetic flux. Fig. 4 may illustrate more clearly the close interrelationship between proximity effect and percolation behavior.

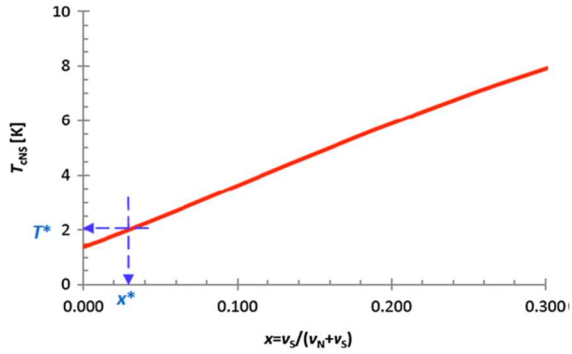

Fig. 2: Critical temperature of the Nb/NbO composite in the Cooper limit of the proximity effect vs the volume fraction x = $v_S/(v_N+v_S)$ of the S component (Nb).

From experiment [6], the temperature dependent part $R_{s,fdt}$ of the field dependent surface resistance increases abruptly at $T^*$ = 2 K (Fig. 3), which indicates percolation behavior.

Indeed, from Fig. 2, the corresponding volume ratio $x^*$ = 0.03 is known as a "void percolation threshold" for "continuum percolation" for a distribution of overlapping spheres (N) with equal radius and voids (S) in between [9].

There, $x^*$ is interpreted as a percolation threshold, and $T^*$ as the corresponding "percolation temperature". They depend on the occupation probability x, where long-range connectivity in random systems (percolation) first occurs.

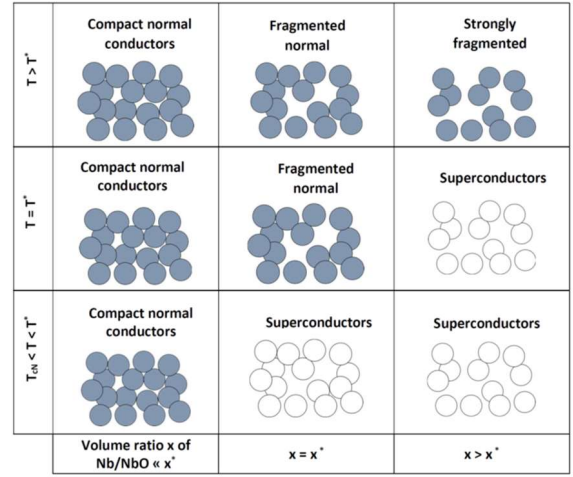

Fig. 4: Scheme of NbO/Nb composites (full and open circles) embedded in a Nb matrix (squares): full circles mean nc; open ones mean sc. Horizontally, the relative volume of the S superconductor is given from below, to just at, and to above the percolation threshold $x^*$. The temperature increases in ascending vertical order, from just above the critical temperature $T_{cN}$ of NbO, but still below the "percolation temperature" $T^*$, to the "percolation temperature" $T^*$, and beyond. The first continuous path from below to above through the Nb matrix and composites appears at $x^*$ and $T^*$. This is equivalent to the start of fragmentation of the composites into small nc defects.

## 4. Nitrogen doping

Doping with nitrogen may improve (or sometimes reduce) the Q value with respect to the field [10]. A binary uniform metal mixture of "dirty" niobium enriched with dissolved nitrogen, as a "weak" superconductor, is subject to the proximity effect produced by the neighbouring high quality niobium metal as a "strong" superconductor. Because of this proximity, the weak superconductor has a lower critical field $B^*$ and lower critical temperature than the strong superconductor. For a very small rf field B, the Q value remains constant until $B^*$ (about 10 - 20 mT), at which its outermost surface candidates become nc. As B continues to increase, the nc zone (volume fraction $f_v$) penetrates deeper into the surface until the field $B_c^*$ (saturation field about 80 - 90 mT) from where on the weak superconductors are fully nc and the Q value remains constant. The weak sc zones with electrical conductivity $s_{Nb}$ are gradually replaced by those with averaged electrical conductivity $s_m$. Since the RF field B at the surface decreases exponentially within the depth x of the superconductor, $B(x) = B \cdot e^{-x/\lambda}$, the RF field at x, B(x), is in turn determined by the logarithm of B, $x = \lambda \cdot \ln(B/B(x))$, $\lambda$ being the penetration depth. Thus, the defected layer ranges from $x = 0$ to $d_N = \lambda \cdot \ln(B_c^*/B^*)$. The function f(B) describes the fraction of the defect layer as a function of the rf field with the range of values between 0 and 1 [11]:

$$f_v(B) = \frac{\ln(B/B^*)}{\ln(B_c^*/B^*)}. \qquad (2)$$

Hence the surface resistance decreases (or increases), depending on $c = s_m/s_{Nb}$, according to the following formula

$$R_s = R_{BCS} \cdot [1 - f_v(B) + c \cdot f_v(B)] \qquad (3)$$

The full amplitude of the magnetic field dependent part determines the constant c (all symbols are explained in Table A-1).

It is interesting to note that the characteristic increase or decrease of the Q value at low fields is also observed in undoped cavities, indicating a rather general phenomenon.

To analyse the data, one turns to a model by R. Landauer, who studied current flow in binary mixtures of media of different electrical conductivities [12]. In this Effective Medium Approximation (EMA) model, the global electrical conductivity $s_m$ is

$$4s_m = (3x_1 - 1)s_1 + (3x_2 - 1)s_2 + \sqrt{[(3x_1 - 1)s_1 + (3x_2 - 1)s_2]^2 + 8s_1 s_2} \quad , \quad (4)$$

where $x_1$ stands for the fraction of the total volume occupied by material 1 (weak superconductor), $x_2$ stands for the fraction of the total volume occupied by material 2 (strong superconductor), and $s_1$ and $s_2$ are the respective electrical conductivities, and $s_m$ is the electrical conductivity of the composite.

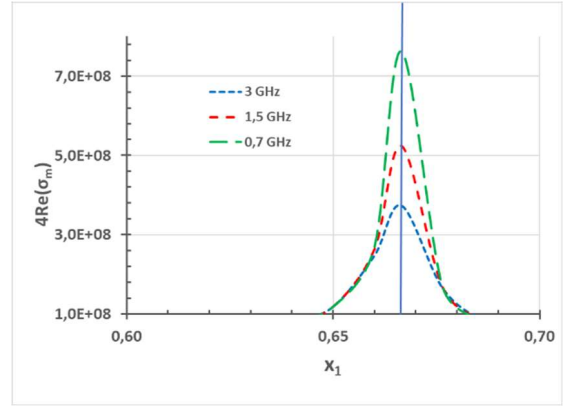

Fig. 5: The real part of the average electrical conductivity $Re(4s_m)$ of the composite at 3 different frequencies; the horizontal axis is the fraction of the total volume $x_1$ occupied by the weak superconductor; $s_1 = 1 \cdot 10^7$ $(\Omega m)^{-1}$, $\lambda$ = 40 nm.

In the present case, the weak superconductor has a typical electrical conductivity of dirty niobium $s_1$, from about $10^6$ to $10^7$ $(\Omega m)^{-1}$. On the other hand, the strong superconductor is pure niobium, of which the electrical conductivity is set to be purely imaginary, $s_2 = i/(\mu_0 \cdot \lambda^2 \cdot \omega)$. Inserting $s_1$ and $s_2$ in eq. 4 results in a curve for $s_m$, the real part of which culminates in a maximum at $x_1 = 2/3$ (Fig. 5). This indicates a percolation path inside the composite. The similarity of this maximum to a (frequency-dependent) bandpass curve of a rf circuit is not accidental; because at the point $x_1 = 2/3$, the real and imaginary parts of $s_m$ are identical, just as in a resonance. This property reflects the assumption of the Landauer model that charges are deposited at defects in a homogeneous medium, much like an LC circuitry.

The often-observed Q-increase (sometimes Q-decrease) at very small fields in niobium bulk cavities is supposed to have the same origin caused by the contamination surface layer which is thinner than that of the nitrogen doped one. This assertion might replace the explanation for the Q increase at low field (due to latent heat) given in ref. 6.

Data as obtained at 2 K and different frequencies [13] could be reproduced making use of the eqs. 1 and 3 in ref. 11 (Fig. 6, continuous lines). Note that these curves are reproduced by only two fit parameters, provided that $x_1 = 2/3$ and the field strength dependence starts at $E_{acc}$ = 5 MV/m as in Fig. 6 (this corresponds to $B^* = 20$ mT).

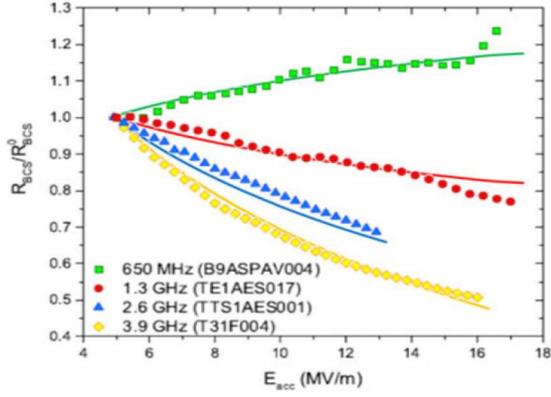

Fig. 6. Surface resistance $R_s$ (here called $R_{BCS}$) at four different frequencies, normalized to the low field surface resistance $R_s^0$ (B = 0) at 5 MV/m for N-doped cavities (made with the exact same doping recipe) as a function of the accelerating field $E_{acc}$ at 2 K; n.b. $B/E_{acc}$ = 4 mT/(MV/m); data adopted from ref. 13.

## 5. Ambient magnetic field

The Q-value depends on the ambient static magnetic field $B_{ext}$, being trapped during cooldown into nc fluxons, as observed in sc cavities of all kinds, although mitigated by fast cooling methods [14, 15]. In the following, data on Nb/Cu cavities are analysed [16]. According to the experimental data of ref. 16 the affected surface resistance $R_{fl}$ can be parametrized as follows,

$$R_{fl} = (R_{fl}^0 + R_{fl}^1 \cdot B) \cdot B_{ext}. \qquad (5)$$

The first contribution can be described as [17]

$$R_{fl}^0 = c_{eff} \cdot (\omega \mu_0)^{3/2} \cdot (2 s_{Nb})^{1/2} \cdot \lambda^2 \cdot \frac{1}{B_{c2}} \;, \quad (6)$$

and is shown in Fig. 7.

The correction factor $c_{eff}$ in eq. 6 (62.5%) takes into account the ratio of the effective magnetic flux component perpendicular to the cavity surface with regard to the overall magnetic flux across the cavity silhouette.

The second contribution is

$$R_{fl}^1 = \frac{1}{s_{Nb} \cdot \lambda \cdot B \cdot B_{c2}} \qquad , \qquad (7)$$

for which a distinction as to the mean free path l has to be made.

For $l \gg \lambda$, the replacements typical of the anomalous skin effect, $s_{Nb} \to s_{eff}$ and $\lambda \to \delta_{eff}$,

$$s_{eff} = \left(\frac{2}{\mu_0 \omega}\right)^{1/3} \left(\frac{\alpha s_n}{l}\right)^{2/3}$$

$$\delta_{eff} = \left(\frac{2l}{\alpha \mu_0 s_{Nb} \omega}\right)^{1/3}$$

lead to

$$R_{fl}^1 \sim \frac{\omega^{2/3}}{\left[\left(\frac{\lambda}{\lambda_L}\right)^2 - 1\right]^{1/3}} \cdot \frac{1}{B \cdot B_{c2}}, l \gg \lambda \qquad . \qquad (8)$$

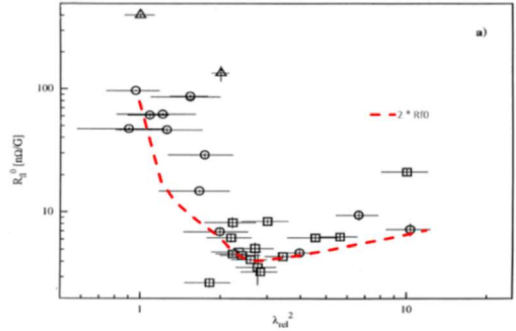

Fig. 7: Trapped fluxon sensitivity $R_{fl}^0$ versus the square of the relative penetration depth $\lambda_{rel} = (\lambda/\lambda_L)^2$. The dashed line represents eq. 6 and is superimposed on the data from ref. 16.

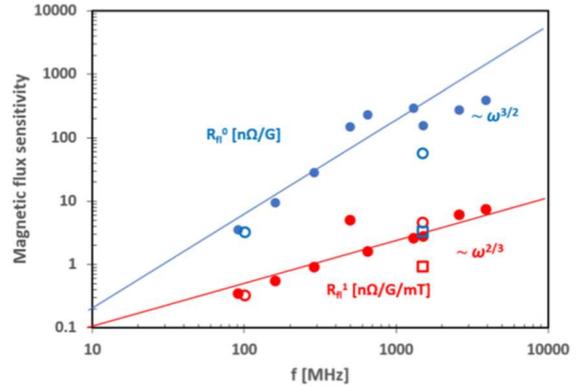

Fig. 8: Frequency dependence of the fluxon sensitivities $R_{fl}^0$ and $R_{fl}^1$ (full dots: bulk niobium; open dots: niobium film). Their dependence on the frequency suggests $\omega^{3/2}$ and $\omega^{2/3}$, respectively (the open squares fall out of the data collection and represent niobium film on oxidized copper cavities, known to have lower values $R_{fl}^0$ and $R_{fl}^1$).

Eq. 8 shows the typical dependence on frequency of the anomalous skin effect and thus confirms the data of Fig. 8.

For $l \ll \lambda$, one obtains

$$R_{fl}^1 = \frac{1}{nel} \cdot \frac{1}{B_{c2}} \sim \lambda_{rel}^2 - 1 \;, \quad l \ll \lambda. \qquad (9)$$

The results from the relations eqs. 8 and 9 are combined and depicted in Fig. 9 [16].

The preceding analysis shows that the observed rf losses in thin film cavities by the ambient magnetic field by trapping can be best described by the following.

1) They originate from fluxons with a local critical temperature around $T_c$ = 4.5 K and a reduced electron density (~17%), compared to standard niobium;

2) they are localized inside and in the close vicinity of these fluxons;

3) they are created by the moving fluxons and the local Hall field directed perpendicular to the current-carrying surface;

4) they follow the anomalous skin effect (for mean free paths larger than the penetration depth) due to the ineffectiveness concept of the shielding current along the fluxons.

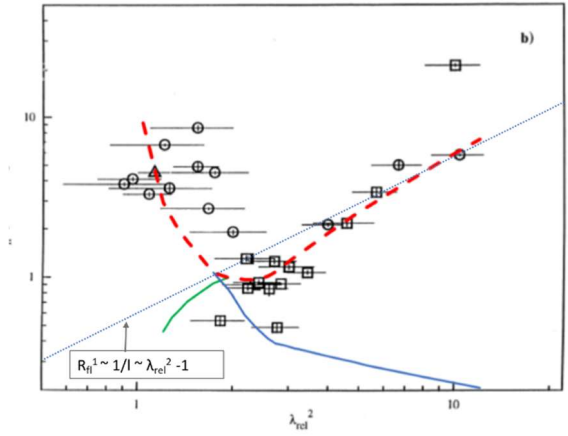

Fig. 9: Trapped fluxon sensitivity $R_{fl}^1$ versus the square of the relative penetration depth $\lambda_{rel}^2 = (\lambda/\lambda_L)^2$. The dashed line is the combination of the two contributions, eqs. 8 and 9, as and is superimposed on the data from ref. 16.

## 6. Ambient magnetic field under N-doping

Nitrogen-doped cavities may respond to an ambient magnetic field by showing a maximum of the trapped fluxon sensitivity at a characteristic mean free path, Fig. 10 [18]. Evidently, the model as explained in section 4 must not be applied here, because the fluxons are fully nc independently of the rf magnetic field amplitude.

This maximum can be explained by the fact that weak sc defects have a different surface resistance than the rest of the niobium surface and therefore build up a space charge when current flows through them. Thus, they act like a capacitor in an alternating field. Together with the inductance formed by the superconductor, they thus form an LC resonant circuit. A lumped-circuit model is used to determine the associated resonant frequency, which, with the help of a fit routine, leads to new results about the properties of the weak sc defects. Compared to standard niobium, they exhibit lower critical temperature and electron density, indicating dirty and/or disordered niobium with many dislocations or dissolved oxygen near the solubility limit.

The fluxon sensitivity $R_{fl}^0$ can be described by [17],

$$R_{fl}^0 = c_{eff} \cdot \frac{R}{1+\left[\left(\frac{\omega}{\omega_0} - \frac{\omega_0}{\omega}\right)\cdot Q\right]^2} \cdot \frac{1}{B_{c2}} \qquad . \qquad (10)$$

The corresponding fit parameters for Fig. 10 are listed in Table 1, the percentage of effective trapped magnetic flux is estimated to $c_{eff}$ = 62.5 % [16], and the other quantities are explained elsewhere [17, Table 1 and Appendix Table A-1].

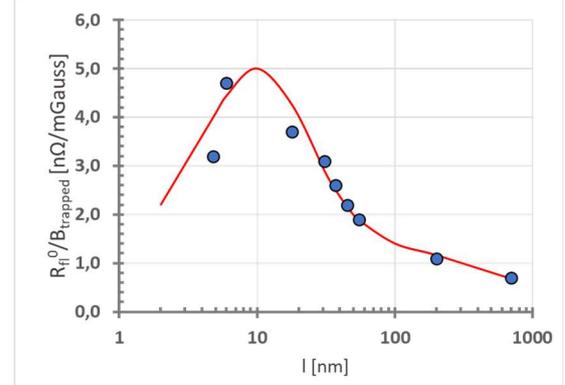

Fig. 10: Trapped fluxon sensitivity $R_{fl}^0$ for N-doped niobium cavities at 1.3 GHz (full dots adopted from ref. 18). The superimposed dashed line (red) results from eq. 10.

Table 1: Fit parameters as to Fig. 10.

| Variable fit parameters | |
|---|---|
| n [m$^{-3}$] | 5.4·10$^{27}$ |
| $s_{Nb}$ (=1/ρ) [(Ωm)$^{-1}$] | 3.3·10$^6$ |
| $\lambda_0$ [nm] | 115 |
| $\xi_0$ [nm] | 49 |
| $B_{c2}(0)$ [Gauss] | 10$^4$ |
| l [nm] /RRR | 3.7 |
| n.b. | |
| $\lambda(l) = \lambda_0 \cdot \sqrt{1+\pi\xi/(2\cdot l)}$ | |
| $\xi(l) = 1/(l^{-1}+ \xi_0^{-1})$ | |

The fit parameters are of the same order of magnitude (within < ±30 %) as those in ref. 17, except $s_{Nb}$ (within < ±80 %), and indicate thus the error margin.

## 7. Role of mean free path under N-doping

The applicability of the model of ref. 11 was also investigated for variable mean free paths l [19]. A contradiction with the experimental data was found. For this reason, the model was discarded as inapplicable [20]. In the following, it will be shown why this criticism is unjustified and must therefore be rejected.

The cause of the discrepancy is a linear relationship, mentioned in ref. 11, between l and the RRR value. This relation is valid for l that is larger than the coherence length ξ. For small l ≈ ξ and below, this linear relation breaks down. However, the authors of ref. 20 applied this relation also for small l, which leads to inconsistencies with the data as shown in Fig. 11.

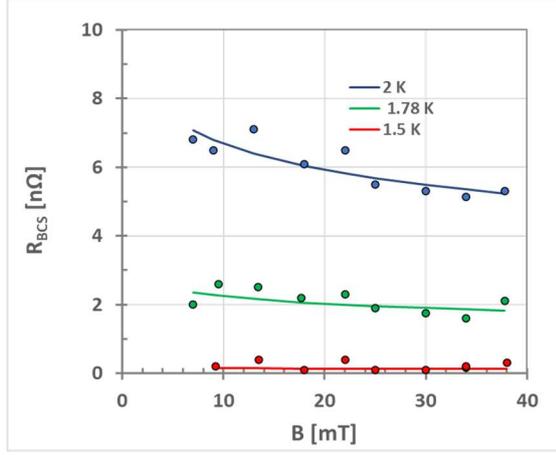

(a)

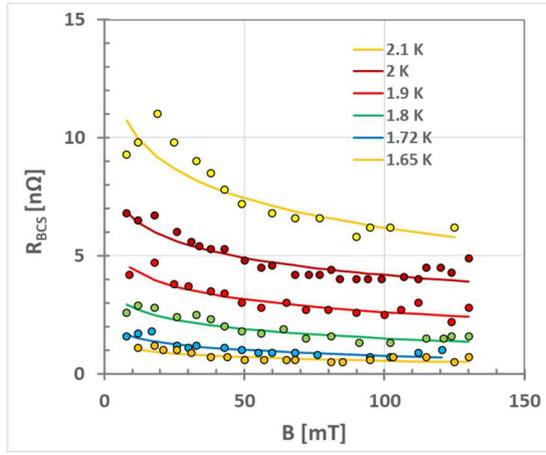

(b)

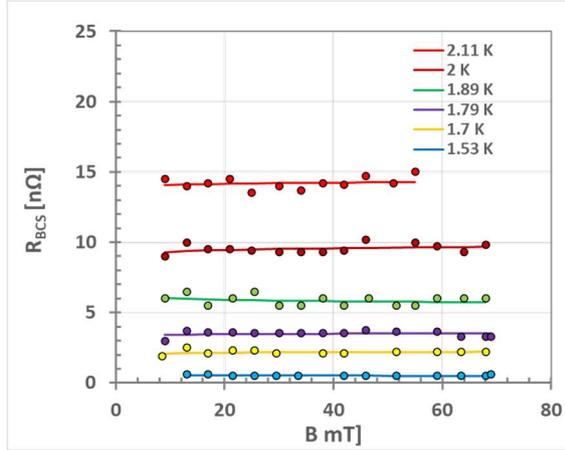

(c)

Fig. 11: Surface resistance $R_{BCS}$ vs. peak magnetic surface field B for 1.3 GHz niobium sheet cavity at different temperatures (in the same vertical order as indicated in the insert). The dots represent the data measured at Cornell University [19], the lines indicate the fitting results with the parameters as in Table 2. The three plots were obtained for three different mean free paths l = 4.5, 34 and 213 nm (from a to c).

To resolve this apparent contradiction, the data of Fig. 11 were analyzed by applying the methods described in section 4; details to be looked up elsewhere [21]. The two parameters c and $s_{Nb}$, which occur in eqs. 1.a and 3, are now determined by means of fitting routines, each for the data of a certain mean free path. The other parameters $B^*$ and $B_c^*$ are obtained by inspection. Thus, $s_m$ is also determined $s_m = c \cdot s_{Nb}$. From the reasonable assumption that the surface resistance is dominated by $s_m$ at the percolation maximum (e.g., Fig. 5), the electrical conductivity $s_1$ (or the electrical resistivity $\rho_1$) of the composite's weak constituent is found for that specific $s_m$ (Table 2, two last columns).

As a consistence check, the RRR value associated with $s_1$ is used to determine the nitrogen concentration. DeSorbo finds for 0.23 (0.33, 1.64) at. % nitrogen interstitially dissolved in niobium a low temperature electrical resistivity of 1.7 (1.9, 1.8) μΩcm [22]. Padamsee gives an RRR value of 3900 for 1 wt. ppm nitrogen [23]. These numbers result in an electrical conductivity $s_1$ (or the electrical resistivity $\rho_1$), Table 3, last column. The two numbers for $\rho_1$ in Table 2 and Table 3 are identical within the error margins, which means that there is no contradiction in the method chosen.

## 8. Ginzburg-Landau analysis as consistency check

The breakdown magnetic fields were studied related to the proximity between a "weak" superconductor (N), being nc if standing alone, and a strong superconductor (S) [24].

The governing quantities in relation with the influence of the magnetic fields on N in close proximity to S are the coherence length $K^{-1}$ and the penetration depth $\lambda_0$ [25]. They have been calculated to $K^{-1} = 236$ nm and $\lambda_0 = 51$ nm, and so was the lower critical magnetic field $B_c \approx 24$ mT [11].

To check these numbers, the general Ginzburg-Landau equations were used. Their range of application is close to the critical temperature, which is considered to be satisfied for the weak superconducting defects discussed here. The cubic term is cancelled, which is justified for the small energy gap Δ in N [24]. These equations are

$$\frac{1}{\kappa^2}\frac{\partial^2 f}{\partial x^2} - f(1+a^2) = 0 \quad , \quad (11)$$

$$\frac{\partial^2 a}{\partial x^2} - f^2 a = 0 \quad , \quad (12)$$

with the reduced energy gap f = Δ/Δ(x=0, B=0), the reduced vector potential $a_0(B) = 2\pi \cdot B \cdot \lambda_0/(\Phi_0 \cdot K)$ and the coordinate x measured from the N/S interface in units of $\lambda_0$.

Table 2: Calculation of the electrical conductivities $s_m$ and $s_1$

| l [nm] | c (average) | $s_{Nb}$ [(Ωm)$^{-1}$] | $s_m$ [(Ωm)$^{-1}$] | $s_1$ [(Ωm)$^{-1}$] | $\rho_1$[μΩcm] |
|---|---|---|---|---|---|
| 4.5 | 0.68 ± 0.03 | (3.1 ± 1.4)·10$^8$ | 2.09·10$^8$ | 4.85·10$^6$ | **21 ± 15** |
| 34 | 0.57 ± 0.05 | (4.9 ± 0.3)·10$^8$ | 2.83·10$^8$ | 7.19·10$^6$ | **14 ± 3** |
| 213 | 1.00 ± 0.04 | (1.03 ± 0.06)·10$^9$ | 1.03·10$^9$ | 7.02·10$^7$ | **1.4 ± 0.2** |

Table 3: Determination of the electrical resistivity $\rho_1$ of "weak" component[1])

| l [nm] | RRR | wt. % N | at. % N | $\rho_1$[μΩcm] |
|---|---|---|---|---|
| 4.5 | 0.64 | 0.61 | 4.0 | **23** |
| 34 | 0.95 | 0.41 | 2.7 | **16** |
| 213 | 9,24 | 0.04 | 0.3 | **1.6** |

[1]) The RRR of the "weak" component is defined as $s_1/s_{Nb}$ (300 K); $s_{Nb}$ (300 K) = 7.6·10$^6$ [(Ωm)$^{-1}$]; $s_1$ from Table 2.

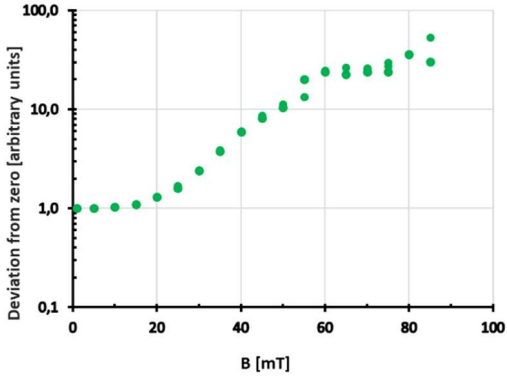

Fig. 12: Mean chi square deviation $\chi^2$ as a measure how close the right-hand side of eqs. 11 and 12 vanish.

The Ansatz for solving the system of coupled non-linear differential equations 11 and 12 is

$$f(x) = cosh\left[(d_N - x)\sqrt{1 + a^2(x,B)} \cdot \kappa\right] + cosh[d_N \kappa(x)] - cosh\left[d_N\sqrt{1+a^2(x;b)}\cdot \kappa(x)\right] \quad (13) \quad ,$$

and

$$a(x,B) = a_0(B) \cdot e^{f(x)\cdot(x-d_N)} \quad , \quad (14)$$

with κ(x) = λ(x)·K, $d_N$ being the depth of the N-doped layer and $\lambda(x) = \lambda_0 \cdot e^{-Kx}$. The two equations 13 and 14 are solved by mutually inserting one into the other and by adding minute trial corrections f´ to f. A fairly good solution is found for the rf magnetic field B up to B* ≈ 20 mT. From here upwards, the right-hand sides of eqs. 11 and 12 deviate more and more from zero up to $B_c^*$ ≈ 60 mT, where the deviation flattens out to a constant value (Fig. 12).

This behaviour is interpreted as reflecting the penetration of magnetic flux into the weak superconductor between its lower critical field B* and the saturation field $B_c^*$. As a comparison, the previous numbers correspond fairly well with the calculated lower critical field B* = 20 mT and the saturation field $B_c^*$ = 66±5 mT [11].

## 9. Conclusion

A total of seven examples of measurements or more general considerations regarding the rf field dependence of the surface resistance in sc cavities are given, all of which can be traced back to a single explanation: weak sc defects at the surface exposed to the rf field. These, if they stood alone, would be nc at sufficiently low temperature. But by proximity to a strong superconductor like niobium, they also become, though only weakly, sc. Arguments are given that these must be defects at the surface. They can increase their expansion by percolation when, for example, increasing the bath temperature. The phenomena amenable to explanation are quite different. They consist of trapping of magnetic flux during the cooling process, abrupt increase of the surface resistance at a specific temperature, anomalous increase of the surface resistance under the influence of the rf field (Q-slope and Q-drop), decrease (or even increase) of the surface resistance after so-called "N-doping", dependent on the rf frequency, or decrease of the surface resistance for variable mean free path. An application of the Ginzburg-Landau equations to these phenomena concludes the paper.

# Appendix

## A-1: Remarks with regard to eq. 1.c

A weakly superconducting hemispherical surface defect is assumed with a dimension a much smaller than the coherence length ξ and much smaller than the penetration depth λ. Furthermore, let the current carrying zone be divided into a relatively contaminated region very close to the surface with a coherence length ξ « λ and a relatively clean region with a coherence length ξ ≈ λ deeper in the interior.

At the moment of a transition at B = B* of the defect to the normal state, the loss of condensation energy $\Delta E_c = B_c^2 \cdot V_c / (2\mu_0)$ is balanced by the gain of diamagnetic energy $\Delta E_B = B^2 \cdot V_m / (2\mu_0)$, where $V_c$ and $V_m$ are the associated volumes and $B_c$ the critical magnetic field of the niobium: $B_c^2 \cdot V_c = B^2 \cdot V_m$. From this, the critical field B* of the defect is derived as

$$B^* = \sqrt{\frac{V_c}{V_m}} B_c \quad .$$

With the hemispherical volumes near the surface (a « ξ, λ),

$$V_m \approx \frac{2}{3}\pi\lambda^3 \quad \text{and} \quad V_c \approx \frac{2}{3}\pi\xi^3 \quad ,$$

B* can be quite small at the surface,

$$B^* = \left(\frac{\xi}{\lambda}\right)^{3/2} B_c = \frac{B_c}{\kappa^{3/2}} \quad ,$$

but deeper in the interior will be close to $B_c$ (κ is the Ginzburg-Landau parameter).

The equality of the two energies $\Delta E_c$ and $\Delta E_B$ defines the volume of the normal conducting zone $V_c$ as a function of the rf field B. The incremental change $\Delta V_c$ then drives the size of the normal conducting zone from a to a+Δa and is given by

$$\Delta V_c \approx \frac{2 \cdot B \cdot V_m}{B_c^2} \cdot \Delta B + \frac{B^2}{B_c^2} \cdot \Delta V_m \quad .$$

With the ratio of the volumina $\Delta V_m / \Delta V_c \approx (\lambda/\xi)^2 \approx \kappa^2$ (Fig. A-1), which are still considered small compared to λ and ξ, so that

$$\Delta V_c \approx \frac{2 \cdot B \cdot V_m}{B_c^2 - B^2 \kappa^2} \cdot \Delta B \quad .$$

The rf power dissipation per square p is proportional to $p \sim V_c \cdot B^2$, such that $\Delta p \sim 2BV_c \cdot \Delta B + B^2 \cdot \Delta V_c$. The first summand represents the rf power loss associated with the square of the rf field and is therefore of no further interest. Only the second summand describes the rf losses, which increase faster than quadratic with the rf field and will therefore be discussed in more detail here. The power dissipation per square,

$$p = \frac{\sigma_n}{4}\omega^2\lambda^3 B^2 \quad ,$$

leads to

$$\Delta p = \frac{1}{4}\omega^2\lambda^3 B^2 \Delta\sigma_n \quad ,$$

and with

$$\frac{\Delta\sigma_n}{\sigma_n} = \frac{\Delta V_c}{V_c}$$

ends up with

$$\Delta p = \frac{\sigma_n}{4}\omega^2\lambda^3 B^2 \frac{\Delta V_c}{V_c} = \frac{\sigma_n}{2}\omega^2\lambda^3 \underbrace{\frac{V_m}{V_c}}_{\approx \kappa^3} \frac{B^3}{B_c^2 - B^2\kappa^2}\Delta B \quad .$$

The rf power dissipation p per square for a high frequency cycle from the low rf field to the high rf field is proportional to the integral

$$p = \frac{\sigma_n}{2}\omega^2\lambda^3 \underbrace{\frac{V_m}{V_c}}_{\approx \kappa^3} \int_{B^*\to 0}^{B} \frac{B'^3}{B_c^2 - B'^2\kappa^2} dB' \quad ,$$

where B* is set close to zero as explained earlier. Finally, after integration [6] and with

$$R_{s,fd}(B,\omega,T) = \frac{2p}{(B/\mu_0)^2} \quad ,$$

the field dependent rf surface resistance $R_{s,fd}$ is obtained as

$$R_{s,fd}(B,\omega,T) =$$

$$= \sigma_n(T)\omega^2\mu_0^2\lambda^3 \underbrace{\frac{V_m}{V_c}}_{\approx \kappa^3} (-\kappa^{-2}) \cdot \underbrace{\left\{1 + \frac{\ln\left[1-\kappa^2\left(\frac{B}{B_c}\right)^2\right]}{\kappa^2\left(\frac{B}{B_c}\right)^2}\right\}}_{eq.1.c} \quad ,$$

the second factor of which corresponds to equation 1.c.

In a more descriptive way, eq.1.c can be developed into an infinite series as follows,

$$R_{s,fdb} \sim \frac{1}{2}\left(\frac{B}{B_c}\right)^2 + \frac{\kappa^2}{3}\left(\frac{B}{B_c}\right)^4 + \frac{\kappa^4}{4}\left(\frac{B}{B_c}\right)^6 + \cdots \quad .$$

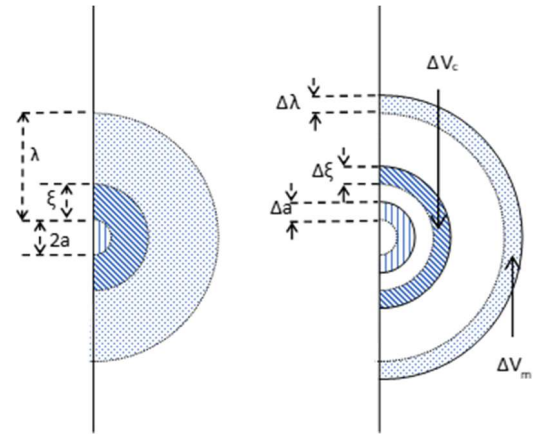

Fig. A-1: Weak superconducting defect at low rf field (left) and after incremental increase of the rf field (right) above the critical temperature B* of the defect.

Table A-1: Description of used symbols

| Symbol | Description | Symbol | Description |
|---|---|---|---|
| a | Reduced vector potential | $K^{-1}$ | Effective coherence length |
| B | Peak rf magnetic surface field | l | Mean free path |
| $B_c$ | Critical magnetic field | n | Density of electrons |
| $B_c^*$ | Saturation field from N-doping | Q | Q-value |
| $B^*$ | Lower critical field for weak superconducting defect | R | Lumped circuit resistance |
| $B_{c2}$ | Upper critical magnetic field | $R_{BCS}$ | "BCS" surface resistance |
| $B_{ext}$ | Ambient magnetic field | $R_{s,fd}$ | rf field dependent surface resistance |
| $c_{eff}$ | Percentage of effective trapped magnetic flux | $R_{s,fdb}$ | Field-dependent surface resistance as a function of the rf field B |
| c | Q-slope improvement (or degradation) ratio from N-doping | $R_{s,fdt}$ | Field-dependent surface resistance as a function of the temperature T |
| $d_N$ | Depth of N-doped layer | $R_{s0}$ | Surface resistance at low field |
| $E_{acc}$ | Accelerating gradient | $R_{fl}$ | Surface resistance from ambient magnetic field $B_{ext}$ |
| f | Reduced energy gap $\Delta/\Delta(x=0, H=0)$ | $R_{fl}^0$ | Component of $R_{fl}$ |
| $f_v$ | Volume fraction of "weak" superconductor | $R_{fl}^1$ | Component of $R_{fl}$ |
| G | Geometry factor | $R_s$ | Surface resistance |

| Symbol | Description | Symbol | Description |
|---|---|---|---|
| $k_B$ | Boltzmann constant | $R_{res}$ | Residual surface resistance |
| $s_1$ | Electrical conductivity of "weak" superconductor | β | Percolation coefficient |
| $s_2$ | Electrical conductivity of "strong" superconductor | $\delta_{eff}$ | Effective penetration depth |
| $S_{Nb}$ | Electrical conductivity of Nb at 10 K | Δ | Energy gap |
| $S_{eff}$ | Effective electrical conductivity | κ | Ginzburg-Landau parameter |
| $s_m$ | Electrical conductivity of composite after transition | λ | Penetration depth |
| $T^*$ | Percolation temperature | $\lambda_L$ | London penetration depth |
| T | Bath temperature | $\lambda_0$ | Penetration depth (x = 0) |
| $T_c$ | Critical temperature of weak sc defect | $\Phi_0$ | Flux quantum |
| $T_{cNS}$ | Critical temperature of the Nb/NbO composite | ρ | Electrical resistivity |
| x | Reduced geometrical length in units of λ | $\omega_0$ | Resonant angular frequency |
| $x_1$ | Fraction of the total volume of weak superconductor | ω | Angular frequency |
| $x_2$ | Fraction of the total volume in strong superconductor | ξ | Coherence length |
| α | Reduction factor for effective electrical conductivity | $\xi_0$ | Intrinsic coherence length (l → ∞) |